\documentclass[a4paper,11pt]{article}
\usepackage{pos}
\usepackage{hyperref}
\usepackage{lineno}
\title{Measurements of inclusive photons at forward rapidities in p--Pb collisions at $\sqrt{s\rm_{NN}}$ = 5.02~TeV with ALICE}
\ShortTitle{Photon multiplicity in p--Pb collisions at $\sqrt{s\rm_{NN}}$ = 5.02~TeV}

\manuallySeparateAuthors

\author*[a]{Abhi Modak}
\author{for the ALICE Collaboration}

\affiliation[a]{Bose Institute, Kolkata, India}

\emailAdd{abhi.modak@cern.ch}

\abstract{We present multiplicity and pseudorapidity distributions of inclusive photons at forward rapidity in proton-lead (p--Pb) collisions at $\sqrt{s\rm_{NN}}$~=~5.02~TeV using the data obtained from Photon Multiplicity Detector (PMD) of ALICE. The centrality dependence of pseudorapidity distributions of inclusive photons is also studied. Results are compared with the previous ALICE measurements of charged-particle production and with theoretical predictions from Monte Carlo models, DPMJET and HIJING.}

\FullConference{%
 
 The Ninth Annual Conference on Large Hadron Collider Physics - LHCP2021\\
 7-12 June 2021\\
 Online
}

\begin{document}
\maketitle

\section{Introduction}
The production of particles in high energy collisions achieved at the Large Hadron Collider (LHC) is dominated by soft-QCD interactions. Soft particle production ($p\rm^{particle}\rm_{T}$ $\lesssim$ 1-2~GeV) is described by non-perturbative QCD and challenges existing phenomenological models. Global observables such as multiplicity and rapidity dependence of particle production are some of the most fundamental measurements for improving and constraining these models. Measurements of these observables in p--Pb collisions provide an important baseline to understand lead-lead (Pb--Pb) results by disentangling cold nuclear matter effects from hot nuclear matter effects~\cite{CNM}. The study of inclusive photon multiplicity aims at providing complementary measurements to those of charged particles since the photon production is dominated by the decay of neutral mesons~\cite{PMDppPaper}.

In this article, the measurements of multiplicity and pseudorapidity distributions of inclusive photons in p--Pb collisions at $\sqrt{s\rm_{NN}}$~=~5.02~TeV are reported. The pseudorapidity distribution (d$N\rm_{\gamma}$/d$\eta\rm_{lab}$) is also measured for different centrality classes from 0--5\% (most central) to 80--100\% (most peripheral). Results are compared with the predictions from DPMJET~\cite{DPMJET} and HIJING~\cite{HIJING} and with the previous measurements of charged-particle production by ALICE~\cite{ChPrpPb,ChPrSysSize}.

\section{Data sample, analysis method, and systematic uncertainties}

This analysis is performed using the ALICE data collected in 2013 by colliding a proton beam of $E = 4~\mathrm{TeV}$ (circulating in the negative z-direction with respect to the ALICE laboratory system~\cite{ALICECordinate}) with a lead beam of $E = 1.58~\mathrm{TeV}$ per nucleon (circulating in the positive z-direction). The detailed description and performance of the ALICE detector system are discussed in Ref.~\cite{ALICEDet}. ALICE has unique coverage to measure both charged particles and photons at forward rapidity. Charged particles can be measured over a wide pseudorapidity range ($\sim$ 8 units) of $-3.4~\textless~\eta\rm_{lab}~\textless~5.0$ using the Forward Multiplicity Detector (FMD) and the Silicon Pixel Detector (SPD)~\cite{FMDpppaper,FMDPbPbpaper}. On the other hand, photon production can be studied over a kinematic range of $2.3~\textless~\eta\rm_{lab}~\textless~3.9$ using the Photon Multiplicity Detector (PMD)~\cite{PMDppPaper}.

The analysis of charged-particle production using FMD and SPD was performed and studied in detail in Ref.~\cite{ChPrSysSize}. In this article, we focus on the new measurements of inclusive photon production using a preshower technique with the PMD. The PMD is a gaseous detector placed at a distance of 367~cm from the interaction point and it consists of two finely granular planes with a lead converter of thickness 3$X_{0}$ ($X_{0}$ = radiation length) sandwiched in between them~\cite{PMDtdr1,PMDtdr2}.

In the present work, minimum bias (MB) events requiring a signal to be detected in both arrays of forward scintillator detectors (V0A and V0C)~\cite{VODetPerform} are analysed. These events correspond to non-single diffractive (NSD) events. Events with vertex position |z| $\textless$ 10~cm from the nominal interaction point are selected. Various centrality classes are determined by measuring the amplitude of the V0A detector~\cite{VODetPerform,ChPrpPbCent}. We apply suitable photon-hadron discrimination thresholds based on the number of affected cells and the energy deposition in a cluster produced in the preshower plane of the PMD to obtain the photon rich sample, known as $\gamma$-like clusters~\cite{PMDppPaper}.

For MB events, the Bayesian unfolding technique~\cite{BayesUnfold} is used to correct the distributions of $N\rm_{\gamma-like}$ clusters which include several detector effects (detection inefficiency, limited acceptance, finite resolution etc.) and contamination from hadron clusters. In this method, all these effects are taken into account by a response matrix ($R$) which is constructed in terms of the true photon ($N\rm_{\gamma-true}$) multiplicity vs $N\rm_{\gamma-like}$ clusters (measured multiplicity) with the help of Monte Carlo (MC) simulation. Each element ($R_{mt}$) of matrix $R$ corresponds to the probability that an event with true multiplicity \textbf{t} is measured as an event with the measured multiplicity \textbf{m}. A detailed description of the unfolding procedure can be found in Refs.~\cite{PMDppPaper,FMDpppaper,ChargedppPaper}. The correction of d$N\rm_{\gamma}$/d$\eta\rm_{lab}$ for various centrality classes is performed using the Efficiency-Purity method as described in Ref.~\cite{STARpmdpaper}.

The MB results are also corrected for trigger and vertex reconstruction efficiencies in a similar way that was followed in Ref.~\cite{PMDppPaper}. The trigger and vertex reconstruction efficiencies are estimated to be 98.47\% and 99.8\% respectively using HIJING.

Systematic uncertainties from various sources (Material budget in front of PMD, discrimination thresholds, unfolding methods, unfolding parameters) are evaluated using the same technique as described in Ref.~\cite{PMDppPaper}. The total systematic uncertainties are obtained by adding uncertainties from various sources in quadrature and found to vary from 4.4 to 57\% and from 7.37 to 7.4\% for multiplicity and pseudorapidity distributions respectively. The systematic uncertainty on d$N\rm_{\gamma}$/d$\eta\rm_{lab}$ due to the centrality classification amounts to 7.38\% and 7.17\% in centrality ranges (0--5\%) and (80--100\%) respectively.

\section{Results and discussion}
\begin{figure}[h!]%CentDepPartPro
	\centering
	\begin{minipage}[b]{0.48\linewidth}
		\includegraphics[width= 1\textwidth]{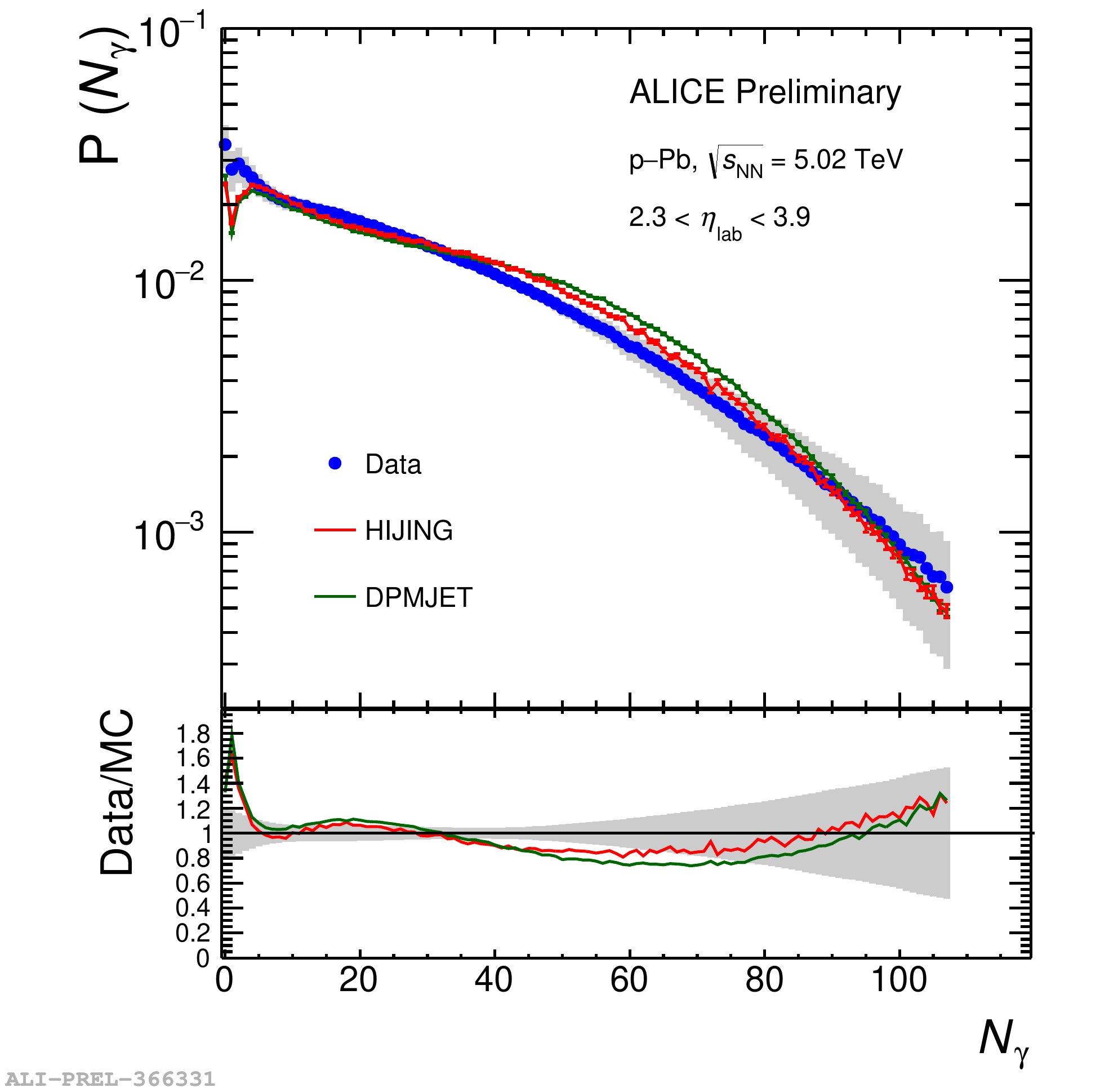}
		\caption{Top panel: Inclusive photon multiplicity distribution for NSD events at forward rapidity, $2.3~\textless~\eta\rm_{lab}~\textless~3.9$, in p--Pb collisions at $\sqrt{s\rm_{NN}}$~=~5.02~TeV. The lines show the predictions from HIJING (Red) and DPMJET (Green). Bottom panel: The ratios between the data and MC results.}
		\label{PNgammaMB}
	\end{minipage}
	\quad
	\begin{minipage}[b]{0.48\linewidth}
		\includegraphics[width= 1\textwidth]{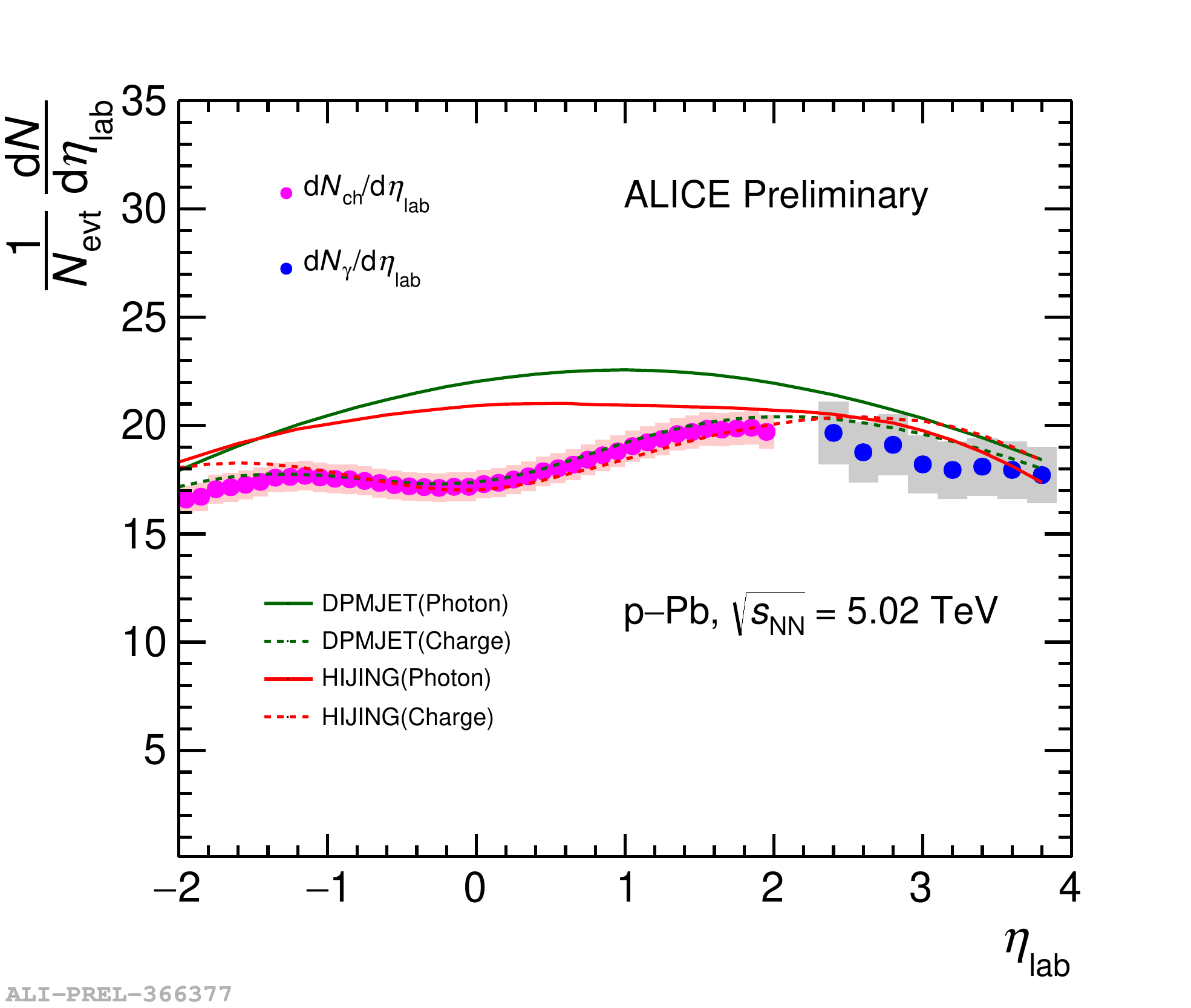}
		\caption{Pseudorapidity distribution of inclusive photons (filled blue circles) for NSD events at forward rapidity, $2.3~\textless~\eta\rm_{lab}~\textless~3.9$, in p--Pb collisions at $\sqrt{s\rm_{NN}}$~=~5.02~TeV and a comparison with previous ALICE measurements of charged-particle production (filled magenta circles) at midrapidity~\cite{ChPrpPb}. Predictions from HIJING (Red line) and DPMJET (Green line) are superimposed.}
		\label{dNdEtaMB}
	\end{minipage}
\end{figure}

Figure~\ref{PNgammaMB} (Top panel) shows the inclusive photon multiplicity distributions measured for NSD events at forward rapidity ( $2.3~\textless~\eta\rm_{lab}~\textless~3.9$) in p--Pb collisions at $\sqrt{s\rm_{NN}}$~=~5.02~TeV in comparison to the predictions from MC models HIJING and DPMJET. The ratios between the data and MC results are shown in the bottom panel of Figure~\ref{PNgammaMB}. It is found that both models underpredict the data at low multiplicities ($N\rm_{\gamma}$~$\textless$~10) and agree with the same in higher multiplicity bins within uncertainties.

Figure~\ref{dNdEtaMB} describes the pseudorapidity distribution of inclusive photons (filled blue circles) as a function of $\eta\rm_{lab}$ within $2.3~\textless~\eta\rm_{lab}~\textless~3.9$ together with previous ALICE results of charged-particle production at midrapidity~\cite{ChPrpPb} (filled magenta circles). Results are compared with the predictions from HIJING and DPMJET. It is observed that d$N\rm_{ch}$/d$\eta\rm_{lab}$ is well described by both MC models whereas d$N\rm_{\gamma}$/d$\eta\rm_{lab}$ is slightly overestimated by DPMJET.

\begin{figure}[h!]
	\includegraphics[scale=0.38]{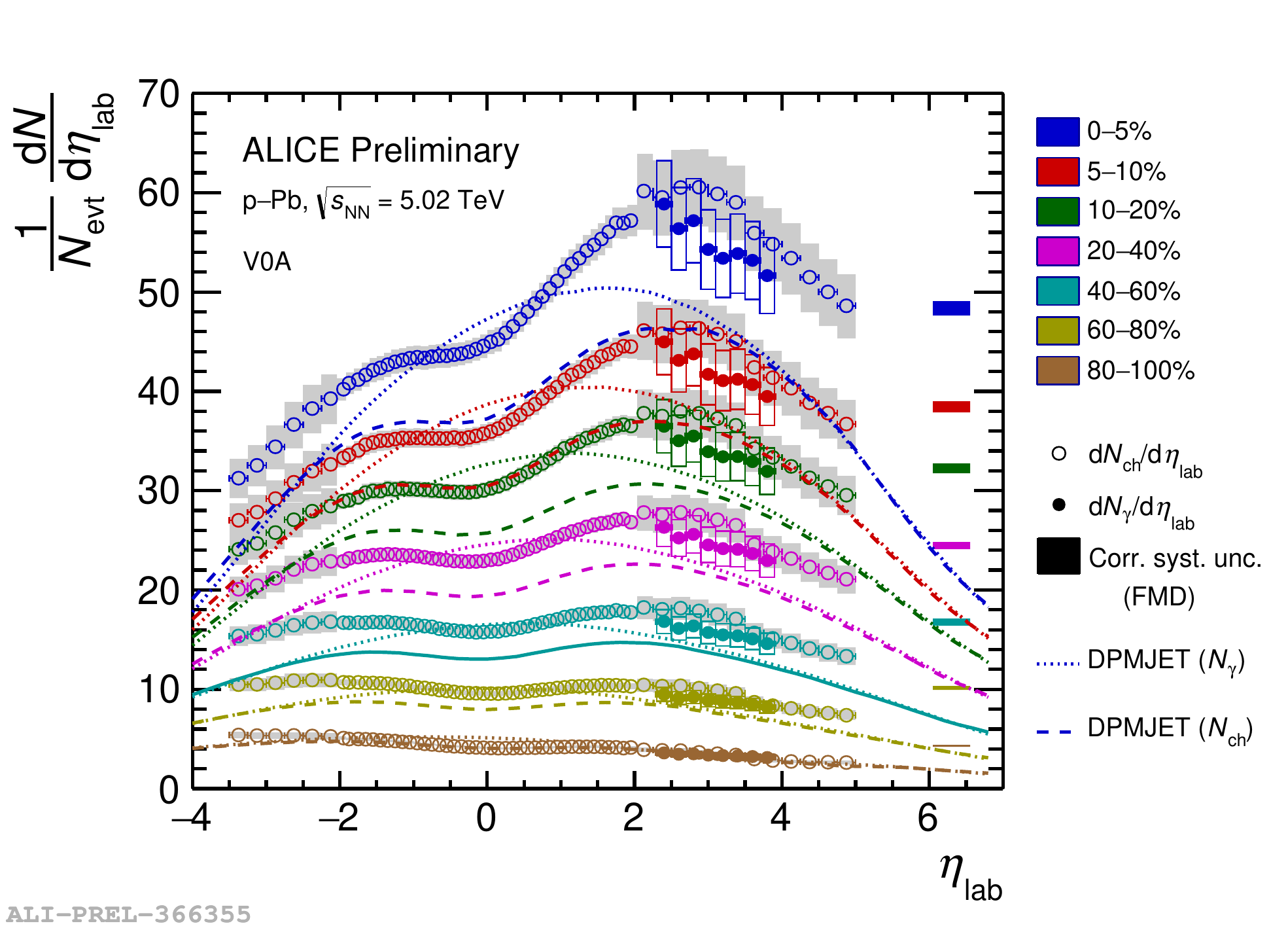}
	\includegraphics[scale=0.38]{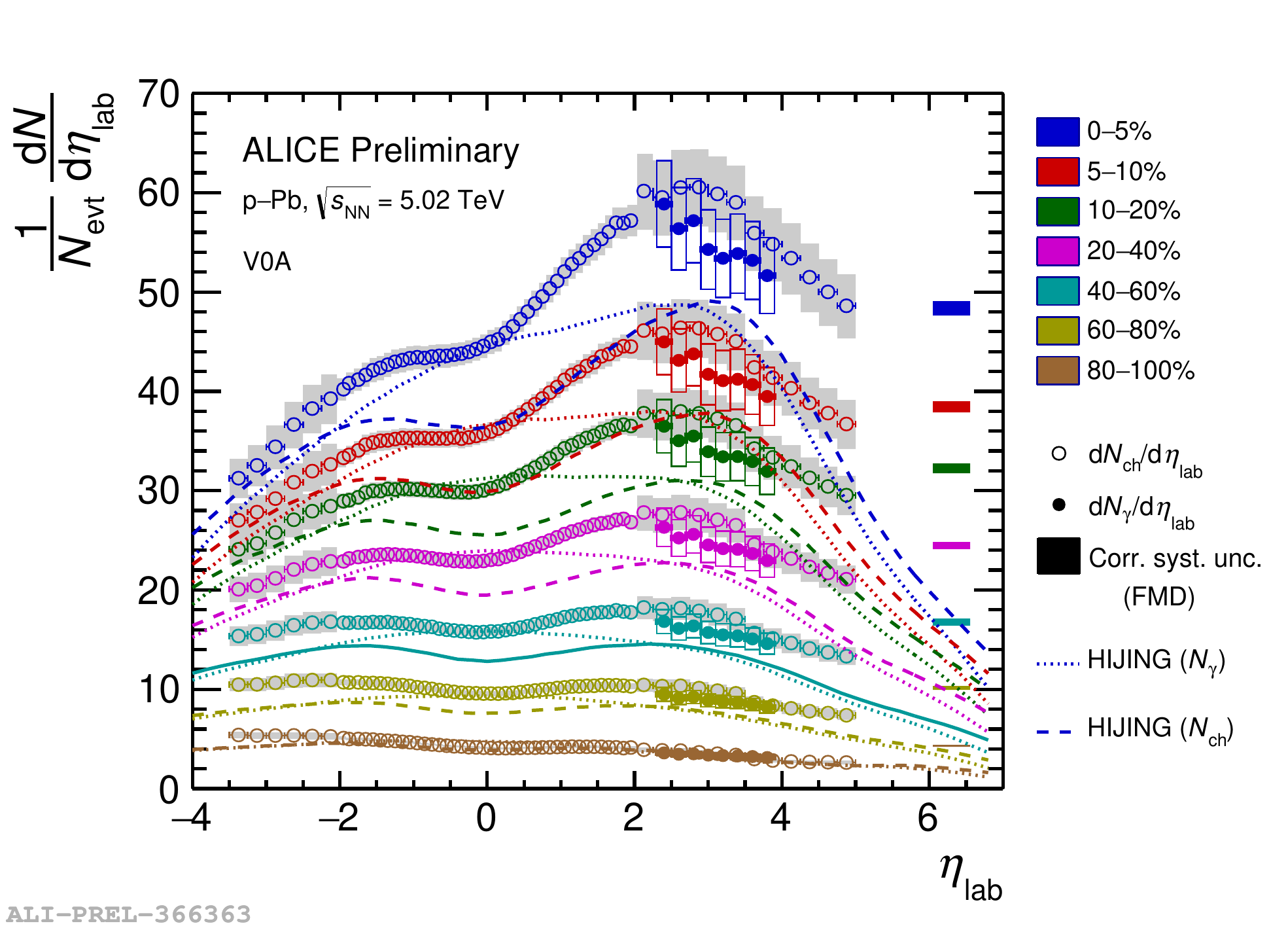}
	\caption{Pseudorapidity distribution of inclusive photons (solid circles) for various centrality classes at forward rapidity, $2.3~\textless~\eta\rm_{lab}~\textless~3.9$, in p--Pb collisions at $\sqrt{s\rm_{NN}}$~=~5.02~TeV and a comparison with previous ALICE measurements of charged-particle production (open circles)~\cite{ChPrSysSize}. Predictions from DPMJET (Left plot) and HIJING (Right plot) are superimposed.}
	\label{CentDepPartPro}
\end{figure}

Centrality evolution of particle production in p--Pb collisions at $\sqrt{s\rm_{NN}}$~=~5.02~TeV is presented in Figure~\ref{CentDepPartPro}. It is seen that inclusive photon (mostly produced from the decay of neutral mesons~\cite{PMDppPaper}) (solid circles) and charged-particle production (open circles)~\cite{ChPrSysSize} have a similar dependence on centrality. A clear asymmetric shape of d$N\rm_{ch}$/d$\eta\rm_{lab}$ is observed for the events with centrality classes from 0--5\% to 20--40\%, with an increasing number of particles produced in the Pb-going direction compared to that in the proton beam direction. Figure~\ref{CentDepPartPro} shows the comparison of the obtained results with the predictions from DPMJET (Left) and HIJING (Right). Both models describe the data for peripheral events (80\%--100\%) however underpredict the same for central events.
 
\section{Summary}
We have presented multiplicity and pseudorapidity distributions of inclusive photons for NSD events in p--Pb collisions at $\sqrt{s\rm_{NN}}$~=~5.02~TeV at forward rapidity ($2.3~\textless~\eta\rm_{lab}~\textless~3.9$) using the data obtained from PMD of ALICE. Results are compared with two theoretical MC predictions, DPMJET and HIJING. We observed that both models underpredict the multiplicity distribution for $N\rm_{\gamma}$~$\textless$~10 and describe the same in higher multiplicity regions within uncertainties. Pseudorapidty distribution of photons is found to be explained by HIJING within uncertainties but slightly overestimated by DPMJET towards midrapidity.

We have also presented photon production for various centrality classes in comparison with previous measurements of charged-particle production by the ALICE~\cite{ChPrSysSize}. A similar dependence of both photon and charged-particle production on centrality classes is observed. Both HIJING and DPMJET are unable to explain the data for all centrality classes except for the most peripheral class (80\%--100\%). These results will help to constrain model parameters to understand the photon and charged-particle productions in p--Pb collisions.


\begin{thebibliography}{99}
{\footnotesize \bibitem{CNM}
J.~Kamin, 
``Hot and cold nuclear matter effects in p\textendash{}Pb collisions at the LHC,''
EPJ Web Conf. \textbf{95}, 03018 (2015)
%doi:10.1051/epjconf/20149503018

\bibitem{PMDppPaper}
B.~B.~Abelev \textit{et al.} [ALICE Collaboration], 
``Inclusive photon production at forward rapidities in proton-proton collisions at $\sqrt{s}$ = 0.9, 2.76 and 7 TeV,''
Eur. Phys. J. C \textbf{75}, no.4, 146 (2015)
%doi:10.1140/epjc/s10052-015-3356-2
%[arXiv:1411.4981 [nucl-ex]].

\bibitem{ChPrSysSize}
C.~H.~Christensen [ALICE Collaboration],
``System-size dependence of the charged-particle pseudorapidity density at $\sqrt {s_{NN}}$ = 5.02 TeV with ALICE,''
Nucl. Phys. A \textbf{967}, 301-304 (2017)
%doi:10.1016/j.nuclphysa.2017.05.066

\bibitem{ChPrpPb}
B.~Abelev \textit{et al.} [ALICE Collaboration],
``Pseudorapidity density of charged particles in $p$ + Pb collisions at $\sqrt{s_{NN}}=5.02$ TeV,''
Phys. Rev. Lett. \textbf{110}, no.3, 032301 (2013)
%doi:10.1103/PhysRevLett.110.032301
%[arXiv:1210.3615 [nucl-ex]].

\bibitem{DPMJET}
S.~Roesler, R.~Engel and J.~Ranft,
``The Monte Carlo event generator DPMJET-III,''
%doi:10.1007/978-3-642-18211-2\_166
[arXiv:hep-ph/0012252 [hep-ph]].
%Ranft J 1994 The Dual parton model at cosmic ray energies Phys. Rev. D 51 64
\bibitem{HIJING}
X.~N.~Wang and M.~Gyulassy,
``HIJING: A Monte Carlo model for multiple jet production in pp, pA and AA collisions,''
Phys. Rev. D \textbf{44}, 3501-3516 (1991)
%doi:10.1103/PhysRevD.44.3501

\bibitem{ALICECordinate}
L.~Betev \textit{et al.},
Definition of the ALICE coordinate system and basic rules for sub-detector components numbering, 
\href{https://edms.cern.ch/ui/#!master/navigator/document?D:1020137949:1020137949:subDocs}{ALICE-INT-2003-038}

\bibitem{ALICEDet}
B.~B.~Abelev \textit{et al.} [ALICE Collaboration],
``Performance of the ALICE Experiment at the CERN LHC,''
Int. J. Mod. Phys. A \textbf{29}, 1430044 (2014)
%doi:10.1142/S0217751X14300440
%[arXiv:1402.4476 [nucl-ex]].

%\cite{ALICE:2017pcy}
\bibitem{FMDpppaper}
S.~Acharya \textit{et al.} [ALICE],
``Charged-particle multiplicity distributions over a wide pseudorapidity range in proton-proton collisions at $\sqrt{s}=$ 0.9, 7, and 8 TeV,''
Eur. Phys. J. C \textbf{77}, no.12, 852 (2017)
%doi:10.1140/epjc/s10052-017-5412-6
%[arXiv:1708.01435 [hep-ex]].
%27 citations counted in INSPIRE as of 04 Sep 2021

%\cite{ALICE:2015bpk}
\bibitem{FMDPbPbpaper}
J.~Adam \textit{et al.} [ALICE Collaboration],
``Centrality evolution of the charged-particle pseudorapidity density over a broad pseudorapidity range in Pb-Pb collisions at $\sqrt{s_{\rm NN}} =$ 2.76 TeV,''
Phys. Lett. B \textbf{754}, 373-385 (2016)
%doi:10.1016/j.physletb.2015.12.082
%[arXiv:1509.07299 [nucl-ex]].
%47 citations counted in INSPIRE as of 04 Sep 2021

%\cite{ALICE:1999hgq}
\bibitem{PMDtdr1}
G.~Dellacasa \textit{et al.} [ALICE Collaboration],
``ALICE technical design report: Photon multiplicity detector (PMD),''
CERN-LHCC-99-32.

\bibitem{PMDtdr2}
P.~Cortese \textit{et al.} [ALICE Collaboration],
``ALICE: Addendum to the technical design report of the Photon Multiplicity Detector (PMD),''
CERN-LHCC-2003-038.

%\cite{ALICE:2013axi}
\bibitem{VODetPerform}
E.~Abbas \textit{et al.} [ALICE],
``Performance of the ALICE VZERO system,''
JINST \textbf{8}, P10016 (2013)
%doi:10.1088/1748-0221/8/10/P10016
%[arXiv:1306.3130 [nucl-ex]].
%250 citations counted in INSPIRE as of 07 Sep 2021

\bibitem{ChPrpPbCent}
J.~Adam \textit{et al.} [ALICE],
``Centrality dependence of particle production in p-Pb collisions at $\sqrt{s_{\rm NN} }$= 5.02 TeV,''
Phys. Rev. C \textbf{91}, no.6, 064905 (2015)
%doi:10.1103/PhysRevC.91.064905
%[arXiv:1412.6828 [nucl-ex]].
%247 citations counted in INSPIRE as of 02 Sep 2021

\bibitem{BayesUnfold}
G.~D'Agostini,
``A Multidimensional unfolding method based on Bayes' theorem,''
Nucl. Instrum. Meth. A \textbf{362}, 487-498 (1995)
%doi:10.1016/0168-9002(95)00274-X

%\cite{ALICE:2015olq}
\bibitem{ChargedppPaper}
J.~Adam \textit{et al.} [ALICE],
``Charged-particle multiplicities in proton\textendash{}proton collisions at $\sqrt{s} = 0.9$ to 8 TeV,''
Eur. Phys. J. C \textbf{77}, no.1, 33 (2017)
%doi:10.1140/epjc/s10052-016-4571-1
%[arXiv:1509.07541 [nucl-ex]].
%111 citations counted in INSPIRE as of 05 Sep 2021

\bibitem{STARpmdpaper}
J.~Adams \textit{et al.} [STAR Collaboration],
``Multiplicity and pseudorapidity distributions of photons in Au + Au collisions at s(NN)**(1/2) = 62.4-GeV,''
Phys. Rev. Lett. \textbf{95}, 062301 (2005)
%doi:10.1103/PhysRevLett.95.062301
%[arXiv:nucl-ex/0502008 [nucl-ex]].
}

\end{thebibliography}
\end{document}